# Characterization of Graphene/Ionic Liquid Memristive Devices for Neuromorphic Systems

*Itır Köymen*[*,1], *Shuyu Liu*[2], *Said Ergöktaş*[2], *Coskun Kocabas*[2]

[1] *Department of Electrical and Electronics Engineering, TOBB University of Economics and Technology, Ankara 06560, Turkey*

[2] *Department of Materials, National Graphene Institute, University of Manchester, M13 9PL, Manchester, United Kingdom*

**ABSTRACT**

Flexible and biocompatible memristive devices are particularly attractive for bioelectronic systems due to the interest in improving computing capabilities and the motivation to interface electronics with biological systems including drug delivery, neural interfaces and biosensors. Structures made of more unorthodox, organic material can address different issues due to their characteristics: flexibility, conformability, biocompatibility and simple and low-cost fabrication. It has been observed that gating graphene/ionic liquid (IL) devices leads to the formation of an electrical double layer (a thin layer of ions with a thickness of a few nanometers) at the graphene/IL interface due to the local potential difference which also controls the local conductivity. This structure provides a memristive mechanism based on a dynamic p-n junction formation along the channel. Motivated by this memristive behavior, graphene/IL devices were assembled with the aim of demonstrating memristive behavior and associative learning. This work investigates memristive properties of flexible graphene/ IL devices on polymer substrates. The I-V characteristics of these novel devices and switching mechanism are investigated. Two distinct topologies (single input, single output and double input, single output) of devices are manufactured and tested to mimic conditioning. It is observed that the application of voltage pulse trains of both positive and negative polarities increases the device conductance and allows larger currents to pass after repetitive excitation. This characteristic was exploited to condition devices and emulate associative learning.

I. Introduction

There has been a shift in the evolution of electronics; while area constraints remain ever present, there is also an interest in improving the functionality of devices. The memristor is a fine example of a device which satisfies both requirements: it is small and has properties which introduce new functionality to electronics such as memory and nonlinear behavior, making it suitable for bioinspired circuits. Following this trend, biocompatible electronic material is increasingly being applied to biomedical applications. The key motivating factor here is that organic material is particularly suitable for interfacing electronics with biological systems including drug delivery, neural interfaces and biosensors [1]. Structures utilizing poly(3,4-ethylenedioxythiophene) doped with poly(styrenesulfonate) (PEDOT: PSS) has allowed taking action potential measurements from the surface of the neocortex and hippocampus without the need for penetrating tissue with electrodes since these structures are conformable. They allow healthy measurements by binding well to the curvilinear surfaces. They are also biocompatible: they do not harm the tissue they contact [2]. A different application utilizes ion gel for creating synaptic transistors for an afferent nerve application. This is demonstrated through the use of the organic material (namely a structures made of styrene-ethylene-butadiene-styrene (SEBS) or polyimide substrates, ion gel and polymer semiconductor, carbon nanotube and Cr/Au layers) as a hybrid system to move a detached cockroach leg. This circuit is also biocompatible and flexible [3]. Motivated by these findings, this work presents graphene/IL devices, characterizes memristive properties and demonstrates learning behavior. ILs can be activated to be sensitive to heat and light. Hence, the combination of graphene/IL is particularly promising for bioinspired applications.

The interest in improving computing capabilities to match the demands of large data has resulted in heightened interest in memory devices [4]. Memristive devices employing a variety of materials are being investigated including transition metal oxides [5], [6], perovskites [7], chalcogenides [8], [9], organic materials [10], [11], [12] and graphene [13], [14], [15], [16], [17]. The more conventional memristor structures (e.g. metal/metal oxide/metal) are fabricated and tested to exhibit certain characteristics such as nanoscale size, stability and reliability. Whereas structures made of more unorthodox, organic material can address different issues due to their characteristics: flexibility, conformability, biocompatibility and simple and low-cost fabrication. The latter type of devices can have diverse topologies, some reportedly have areas in the range of $mm^2$ [18] and even $cm^2$ [3]. The devices fabricated and investigated in the present work benefit from the low-cost, simple fabrication methods.

Graphene/IL devices have been investigated due to their optoelectronic properties [19], [20]. Electrical and optical properties of graphene can be electrostatically tuned [21] [22]. When used with electrolytes or IL, gating graphene/IL devices leads to the formation of an electrical double layer (a thin layer of ions with a thickness of a few nanometers) at the graphene/IL interface [20] due to the local potential difference which also controls the local conductivity as seen in Figure 1 (a). This structure provides a memristive mechanism based on a dynamic p-n junction formation along the channel. At low bias voltages, the p-n junction blocks the current flow through the graphene and yields high resistive state (HRS) as seen in Figure 1 (b). This self-formed p-n junction is under forward bias, therefore increasing the bias voltage lowers the potential barrier and enhances the current. Furthermore, the Dirac point shifts to negative voltages which increases the local doping. Combination of these competing effects yields a threshold voltage. After this threshold voltage, the device enters the low resistive state (LRS) where the condition for p-n junction is broken. As a result of this mechanism, it has been observed that graphene/electrolyte/graphene structures exhibit capacitance and resistance which are strongly dependent on input voltage [20]. This behavior is attributed to gate-induced change of Fermi energy of graphene due to the bias voltage. Maximum resistance (and minimum capacitance) is observed when carrier concentration is at a minimum. This quality is in fact very similar to the ionic movement in memristive devices [5], [23], [24]. It is not surprising that hysteric I-V behavior has already been observed in other similar structures [25] as well as the graphene/IL devices studied in this work, since as with memristors charge carriers move due to an excitation signal and concentrate on one side of the device which leads to a time dependent resistance variation, upon reversing the polarity of the excitation signal, ions move to the opposite side of the device and opposite (incremental or decremental) change in resistance in time is observed.

The graphene/IL devices on flexible polymer substrates presented in this work exhibit I-V responses which are nonlinear and memristive with clear high resistive and low resistive states (HRS and LRS respectively). The devices assembled for the purposes of this work are not of the conventional sandwich-like memristor variety. They have a lateral dynamic doping profile which generates a lateral p-n junction as shown in Figure 1 (d).

Separately from applications utilizing graphene, devices with IL (with or without graphene) are known to behave as self-switching diodes. This dynamic p-n junction type behavior arises when bias is applied to IL devices from one electrode (with the other electrode grounded), anions and cations separate from their homogenous form at 0V bias and negative electrode becomes highly populated with cations and positive electrode attracts the anions. This leads to a shift in the local Fermi level, leading to diodic I-V behavior [26]. Another study shows preliminary results of an IL/water device with Pt electrodes which exhibit hysteresis [12]. These results encouraged the use of IL for memristive devices.

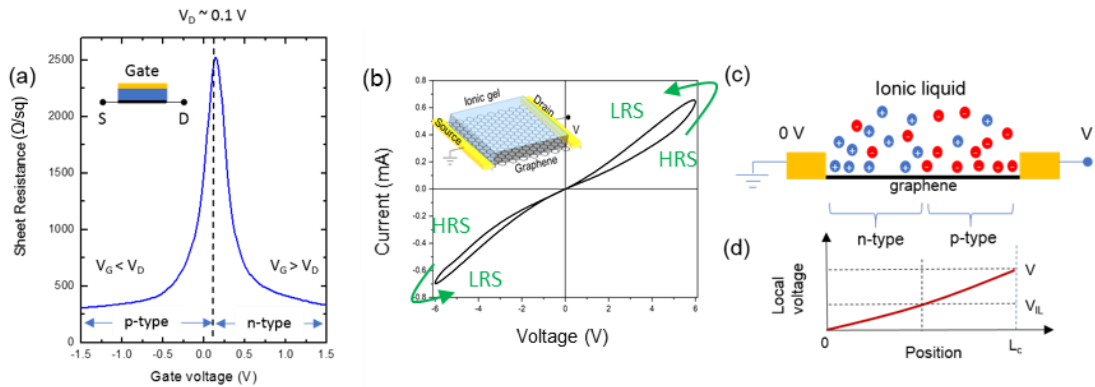

**Figure 1.** Mechanism of ionic gating and hysteresis in graphene and graphene/IL devices (a) Variation of the sheet resistance of graphene with the gate voltage applied to the electrolyte. Graphene shows ambipolar transport for both electrons (VG>VD) and holes (VG<VD). Dirac point is around 0.1V. (b) Graphene/IL memristor:  I-V curve of graphene memristor depicting low (LRS) and high resistance state (HRS) with pinched hysteresis. Device layout is shown in the inset. (c) Schematic representation of a dynamic p-n junction formed on the graphene channel. The variation of the local potential redistributes ions in the electrolyte and forms dynamic doping on graphene (d) Variation of the local potential along the channel under a bias voltage. IL behaves as a floating gate which yields position dependent gating.

Associative learning is a process during which two seemingly independent experiences are mentally linked to one another. A familiar example of associative learning is the Pavlov's dog experiment, where a bell is rung each time Pavlov's dog is given food until finally the dog begins salivating at the sound of the bell even when there it is not presented with food. This learning mechanism whereby the body or brain begins to expect a certain type of stimulus after experiencing an independent stimulus is also called conditioning. This type of learning behavior which can be observed in memristive devices of various forms have been reported [27], [28].

Motivated by the myriad of intriguing qualities (tunable thermal radiation, electrochromic, memristive characteristics) of graphene/IL devices, we assembled two different configurations of such devices with the aim of demonstrating memristive behavior and conditioning. It is important to note that (1) graphene provides ambipolar conduction, (2) graphene based resistor, memristor and transistor [29] can be fabricated on the same device, (3) low voltage operation and dynamic p-n junction yields high cyclic endurance. These features could enable complex neuromorphic circuits that can emulate signal processing and conditional learning.

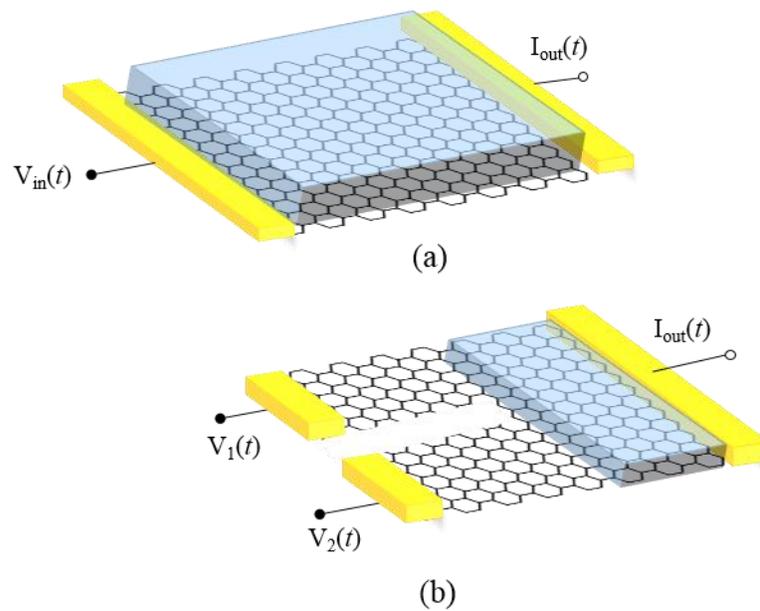

**Figure 2.** Graphene/IL self-gated devices on flexible polymer substrates (a) Single input-single output device (b) Double input-single output device

## II. Device Fabrication

To fabricate the graphene based memristive devices we purchased the commercially available chemical vapor deposition grown single-layer graphene on copper foil from MCK Tech Co. Ltd. 75-μm thick polyethylene terephthalate (PET) film was laminated on graphene to mechanically support the graphene and give flexibility to the final device. Then, copper foil was etched in 0.1M ammonium persulfate (APS) solution and rinsed with DI water following the etching. Electrical contacts were taken *via* conductive carbon tapes. 25-μm thick porous polyethylene (pe) membrane was placed on top of graphene to perform as host medium to the IL electrolyte, [DEME] [TFSI] (Sigma Aldrich, #727679), which dropped into the membrane as a last step of the fabrication.

Electrical characterizations of the graphene/IL devices were performed using National Instrument CompactDAQ system (NI 9264 and NI9203).

## III. Device Characterization

The graphene/IL devices are considerably different compared to majority of devices which are and have been investigated with regards to memristive behavior, and are in the vein of the newer, biocompatible, flexible and low cost organic material based memristive devices [10], [11], [12], [13], [14], [15], [16], [17]. Their structure does not follow the common recipe of metal/oxide/ doped oxide/ metal [5], [6]; therefore, it was important to conduct preliminary measurements to characterize these devices and see whether they exhibited memristance as described in [30]: Pinched hysteresis loops exhibiting passive behavior (1$^{st}$ and 3$^{rd}$ quadrants) which are bound by two distinct resistance values ($R_{on}$- low resistive state and $R_{off}$- high resistive state), varying in size according to the input signal amplitude and frequency were in fact observed. The measurements presented in this work are taken from large samples (area in the order of cm$^2$) as opposed to nano/micrometric devices. Another important reason for these preliminary measurements is to determine the input signal amplitude and frequency range for the experiments to follow. Not having much prior information with regards to where these devices operate, it was crucial to pinpoint frequency and amplitude response to determine pulse lengths and strengths.

Graphene/IL devices were assembled on flexible polymer substrates. I-V measurements were performed by applying ac voltage signal of different frequencies to observe hysteresis and the variation of device behavior due to differing driving frequencies. A drawing of the single-input single-output devices measured in this capacity is depicted in Figure 2 (a).

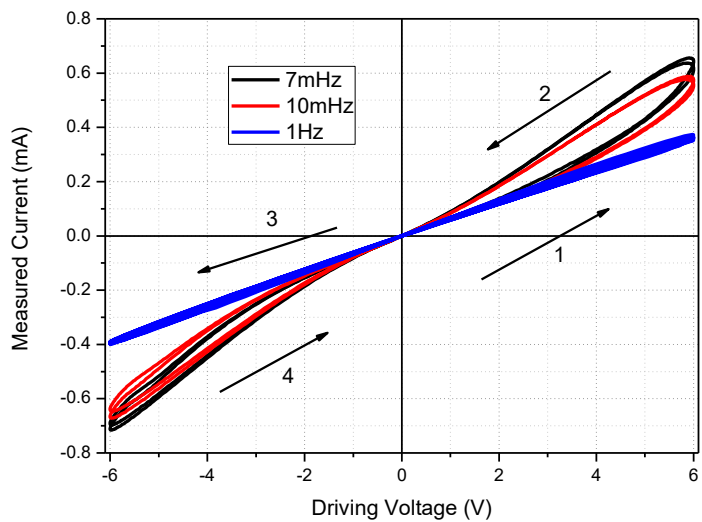

**Figure 3.** Hysteresis plots of Graphene/IL devices at 3 distinct driving signal frequencies. The plots get wider as driving frequency diminishes. The switching direction is denoted with arrows.

Figure 3 depicts the characteristic pinched hysteresis loops for graphene/IL devices. The devices exhibit passive behavior: their I-V characteristics are limited to the 1$^{st}$ and 3$^{rd}$ quadrants. The frequency dependency is also as expected; higher frequencies lead to more linear behavior, hysteresis loops get wider for lower frequencies.

Investigating the hysteresis behavior of graphene/IL devices: it is observed that in both quadrants, the memristance goes from $R_{off}$ to $R_{on}$. We expect this to happen in the first quadrant, the device is turned on due to excitation voltage; hence, its resistance decreases as voltage is applied. The behavior of the memristor reported by Hewlett-Packard (HP) in their seminal paper is also investigated thoroughly in terms of switching direction due to ac excitation voltage [5]. HP's device and the devices studied here share the switching qualities in the first quadrants. However, their behavior in the negative half of the signal differ. When the negative half of the ac signal is applied HP's device, it switches from $R_{on}$ to $R_{off}$. However, in the case of these graphene/IL devices 0V resets the devices and the memristance goes from $R_{off}$ to $R_{on}$ as it did in the 1$^{st}$ quadrant. This demonstrates that the memristance of the graphene/IL devices varies identically for both the positive and negative segments of a bipolar signal. The memristance therefore, does not show a continuous variation over a single period of and ac signal but rather switches back to its off state when the excitation is at 0V. This observation is valuable in terms of understanding IL switching mechanism and for utilizing these devices in more complex systems.

In order to further demonstrate the time dependent nature of the graphene/IL devices, Figure 4 shows the current response to an irregular voltage pulse train. This set of pulses is helpful in highlighting the effect of larger and/or longer pulses on the memristive behavior of the graphene/IL device. The current magnitude increases due to repeated positive pulses seen from the first five narrow pulses. The wider pulses show that current gradually increases when the pulses are held for longer: e.g. second set of pulses are slightly longer than the first and show the increasing gradient, third set have a larger amplitude and are held for slightly longer- current slope is apparent, the fourth pulse set shows that current follows the shape of the voltage pulse until the slightly longer negative pulse where a negative slope is observed due to the time interval at the negative voltage, the fifth pulse set-the longest pulse- clearly shows the positive slope in current as it increases due to the time interval spent at this value. Exciting the graphene/IL devices with irregular pulses demonstrate the time dependent nonlinear behavior of the devices clearly. Holding the voltage amplitude at a certain value for longer periods of time decrease the memristance of the device since this value is dependent on the excitation magnitude and duration, this in turn leads to an increase in current response.

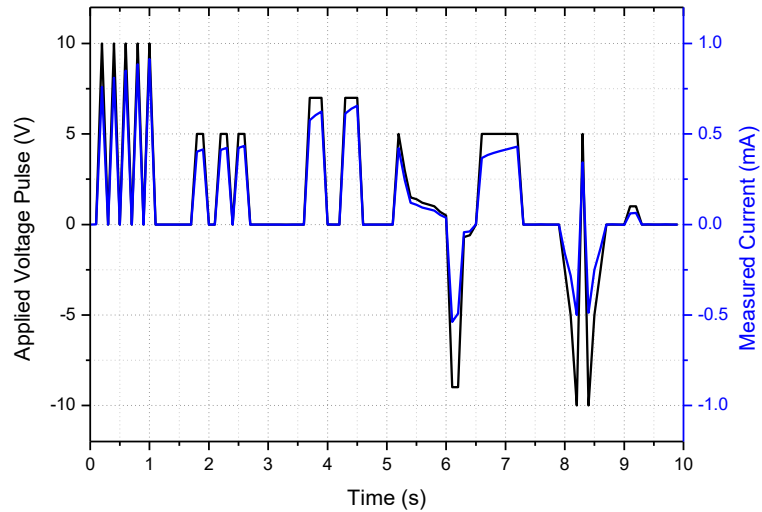

**Figure 4.** Single input flexible polymer substrate graphene/IL device excited by irregular pulses to show effect of longer pulses upon measured current.

### IV. Results and Discussion: Pulse Train Measurements for Learning Mechanism

Figure 5 shows the measurement results recorded after the application of identical pulse trains to flexible polymer substrate graphene/IL devices. What is expected from a typical bipolar memristive device is that its memristance will decrease due to a positive excitation signal (for a voltage driven device this will mean measured current will increase in value due to continuous positive driving voltage) and a negative input signal will lead to an increase in memristance (lowering the resulting measured current value). However, as it was seen in our I-V measurements, the graphene/IL devices exhibit the same switching direction in both positive and negative input cycles. As such, while the measured current response to positive voltage pulses is as expected (a decreasing memristance: HRS to LRS), the response to negative pulses indicate decreasing memristance. The memristance of the devices seems to increase during polarity switches, then decreases for both negative and positive inputs. Absolute value of current magnitude due to repetitive application of both negative and positive pulses increase. This is consistent with the assessment made with reference to the hysteresis plots, once the driving signal is 0V, the device resets. It can be said that the current response is symmetrical.

It is also important to note that conditioning is demonstrated with these simple pulse train experiments. Voltage pulses periodically rise to 10V and fall back down to 0V. Had these devices merely exhibited linear resistance, the measured currents would look like a scaled version of the pulse trains, reaching a certain constant current value for every pulse. However, the nonlinear, time dependent I-V characteristic of the memristive graphene/IL devices is apparent in Figure 5. The excitation is 0V between each pulse as seen in Figure 5; thus, the device resets, current is also 0A. Following the reset after both positive and negative cycles, the magnitude of measured current of the device begins to increase while being applied repeated pulse trains of the same polarity. This behavior is due to the increasing conductance of the graphene/IL devices resulting from being applied repeated pulses. Similar to physiological systems where the increase in channel conductance translates to more being delivered [31] to the post-synaptic neuron , the conductance of this device too increases due to repeated excitation and thus larger current flows through the device.

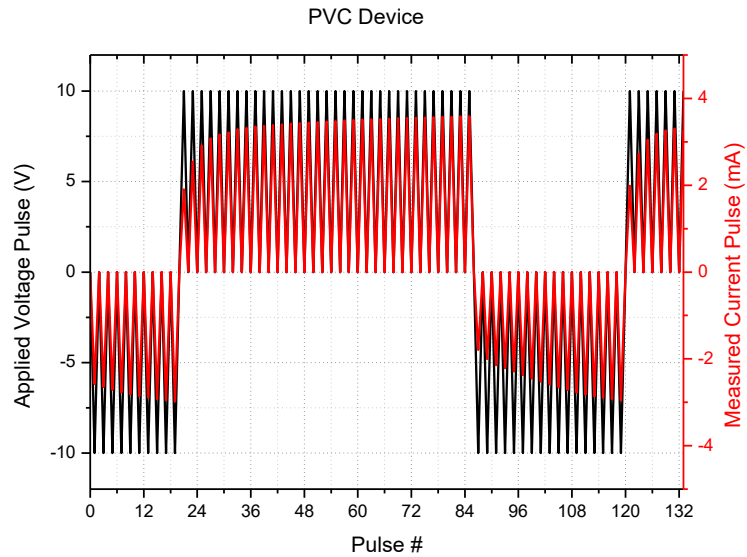

**Figure 5.** Flexible polymer substrate Graphene/IL device excited with repetitive positive and negative 10 V pulse trains. Current response shown in red.

**Two input one output device**

A two input one output graphene/IL device is assembled as shown in Figure 2 (b). This configuration is different to the typical memristor topology; yet similar to multi-gated devices seen in literature [3]. This configuration allows mimicking associative learning. It is also useful in observing the behavior of different geometries of graphene/IL devices.

Each pulse in Figure 6 is 0.1s long. The graphene/IL device is driven with two distinct voltage pulse sources $V_1$ and $V_2$ shown in black and blue in Figure 6 respectively. The resulting current through the device is measured and shown in red. This figure demonstrates three important characteristics:

1. This configuration, like the simple one port device, leads to conductance variation due to applied voltage pulses. This is seen in the measured current shown in red. Repetitive pulses of constant value (10V or 5V in this case) lead to increase in the measured current (increase in conductance, decrease in resistance). This can be observed between Pulse #'s 10-19, 30-37, 38-67, 68-70.  This is similar to how synaptic weight change occurs- repeated positive input, increases synaptic weight, increases the conductance of channels.

2. Applying voltage pulses simultaneously leads to a higher current through the device. When the voltage pulses from the two sources are superimposed, the resulting voltage across the device is higher. Thus, the memristance is lowered more and a higher current flows through the device. Looking at the device overall, this is consistent with the results shown in Figure 5 and Figure 4, increased input voltage leads to increased input current.

3. Perhaps most the exciting result is that we observe behavior that is analogous to associative learning. When the device is driven with $V_1$=-10V alone initially, the measured current is -2.66mA. We then apply $V_1$=$V_2$=-10V simultaneously, current through the device increases as mentioned above. Then when $V_1$=-10V is applied alone again, the measured current is 2.8mA, goes up to 2.84mA before it decreases to 2.78mA. This increase is due to the fact that the device has been conditioned to expect a second pulse simultaneously. Even though, it is only driven by $V_1$, the device behaves as if it is being applied $V_1$ and $V_2$ simultaneously and therefore the current through it is larger than it was when it was originally applied the same amplitude of the single input voltage.

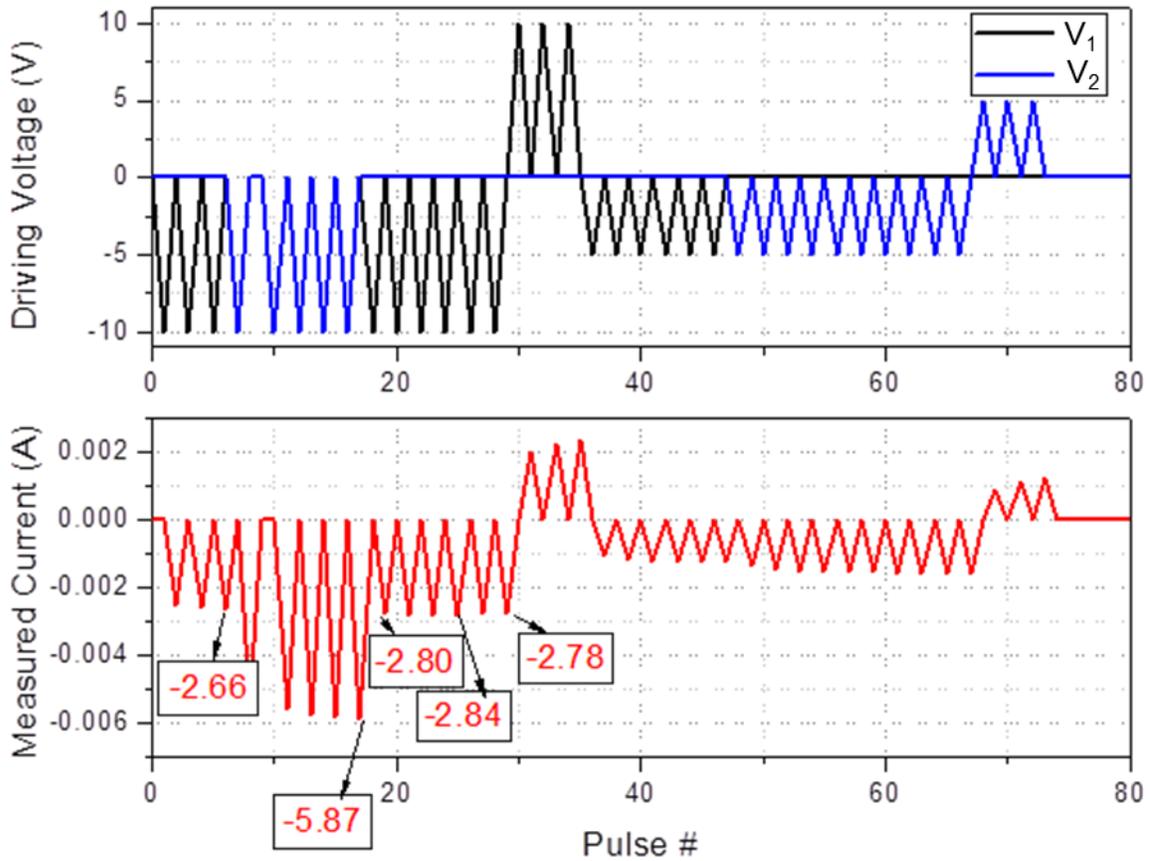

**Figure 6.** The response of a 2-input terminal 1-output terminal flexible polymer substrate graphene/IL device to voltage pulses.

In Pavlov's experiment of conditioning the dog to expect food at the sound of the bell, conditioning stimulus is the bell ringing, in our experiments this can be likened to $V_1$ or $V_2$ being applied separately (while one of the stimuli is at 0V). The unconditioned stimulus in Pavlov's experiment is food. When conditioned and unconditioned stimuli is applied simultaneously, an unconditioned response occurs, in Pavlov's experiments this corresponds to salivation. In this setup, this corresponds to $V_1$ and $V_2$ being applied simultaneously to increase conductance of the device and lead to a large current response. This can also be called the training phase of the experiment. Following this, once the unconditioned stimuli is withdrawn (food), unconditioned response is still achieved due to associative learning. With the two input one output graphene/IL device, we observe a similar result, $V_1$ and $V_2$ are applied simultaneously, a large current passes through the device, $V_2$ is withdrawn, $V_1$ is applied alone, a current value larger than what was recorded before training is observed. The device thus demonstrates characteristics that enable mimicking associative learning.

## V. Conclusions

The switching properties instilled by employing IL in electronic devices were utilized in this work for the purpose of achieving memristive I-V characteristics. IL with graphene, on flexible polymer substrates yield flexible, low-cost devices. We show in this work that these devices exhibit repeatable memristive behavior and that these switching qualities can be harnessed to demonstrate learning. Some key findings particular to these devices are that these devices exhibit pinched hysteresis loops which vary with frequency (higher frequencies lead to more linear behavior, lower frequencies widen hysteresis loops and lead to more nonlinear behavior). Secondly, these graphene/IL devices' conductance responds to positive and negative stimuli in the same way. We have experimented with numerous devices and different topologies. Following characterization measurements showing pinched hysteresis, we demonstrate memory with pulse train measurements: repetitive pulses led to decreasing conductance and gradually increasing current output pulses. We also conducted measurements with irregular pulse trains to also point to the inherent integration property of memristors. The device characteristics show potential for use in learning applications. Wider pulses decreased the conductance of the device and resulted in an increasing current response. The two input terminal-one output terminal device measurements point to associative learning, following simultaneous excitation from both terminals, the conductance decreases, after removing one of the voltages, the conductance increases again but once the single pulse is sent again, the device, expecting the second pulse simultaneously, becomes more conducting again.

The graphene/IL devices presented in this work are novel memristive devices which are different to CMOS compatible, semiconductor memristors. Having characterized these graphene/IL devices to ensure that they are memristive, and conducted experiments to demonstrate learning properties, it can be deduced that these flexible and scalable devices are very promising for further applications where devices can be trained with electrical signals and be made into heat or light sensitive devices to act as compact sensor/ processor units.

## AUTHOR INFORMATION

**Corresponding Author**

* **Itır Köymen** - *Department of Electrical and Electronics Engineering, TOBB University of Economics and Technology, Ankara 06560, Turkey*

Email: ikoymen@etu.edu.tr

## ACKNOWLEDGMENT

Itır Köymen acknowledges the partial support of this work by the Scientific and Technological Research Council of Turkey (TUBITAK) 2218 Scheme. Coşkun Kocabaş acknowledges the support from the European Research Council (ERC) Consolidator Grant ERC-682723 SmartGraphene.

## REFERENCES

[1] P. Gkoupidenis, D. A. Koutsouras and G. G. Malliaras, "Neuromorphic device architectures with global connectivity through electrolyte gating," *Nature Communications,* vol. 8, p. 15448, 2017.

[2] D. Khodagholy, J. N. Gelinas, T. Thesen, O. Devinsky, G. G. Malliaras and G. Buzsáki, "NeuroGrid: recording action potentials from the surface of the brain," *Nature Neuroscience,* vol. 18, pp. 310-315, 2015.

[3] Y. Kim, A. Chortos, W. Xu, Y. Liu, J. Y. Oh, D. Son, J. Kang, A. M. Foudeh, C. Zhu, Y. Lee, S. Niu, J. Liu, R. Pfattner, Z. Bao and T.-W. Lee, "A bioinspired flexible organic artificial afferent nerve," *Science,* vol. 360, no. 6392, pp. 998-1003, 2018.

[4] Z. Shen, C. Zhao, Y. Qi, I. Z. Mitrovic, L. Yang, J. Wen, Y. Huang, P. Li and C. Zhao, "Memristive Non-Volatile Memory Based on Graphene Materials,," *Micromachines,* vol. 11, no. 4, 2020.

[5] D. B. Strukov, G. S. Snider, D. R. Stewart and R. Williams, "The missing memristor found," *Nature,* vol. 453, pp. 80-83, 2008.

[6] B. Mohammad, M. A. Jaoude, V. Kumar, D. M. Al Homouz, H. Abu Nahla, M. Al-Qutayri and N. Christoforo, "State of the art of metal oxide memristor devices,," *Nanotechnology Reviews,* vol. 5, no. 3, pp. 311-329, 2016.

[7] K. Ochkan, S. Razmkhah, P. Febvre, E. Zhitlukhina, Belogolovskii and M, "Investigating a Discrete Model of Memristive Systems," in *XIth International Scientific and Practical Conference on Electronics and Information Technologies (ELIT)*, Lviv, 2019.

[8] Y. Li, Y. Zhong, J. Zhang, X. Xu, H. Sun and X. Miao, "Ultrafast Synaptic Events in a Chalcogenide Memristor," *Scientific Reports,* vol. 3, no. 1, 2013.

[9]  K. A. Campbell, R. A. Bassine, M. F. Kabir and J. Astle, "An Optically Gated Transistor Composed of Amorphous M + Ge2Se3 (M = Cu or Sn) for Accessing and Continuously Programming a Memristor," *ACS Appl. Electron. Mater,* vol. 1, no. 1, pp. 96-194, 2019.

[10] N. Raeis-Hosseini and J. Rho, "Solution-Processed Flexible Biomemristor Based on Gold-Decorated Chitosan," *ACS Appl. Mater. Interfaces,* 2021.

[11] Y.-C. Chen, H.-C. Yu, C.-Y. Huang, W.-L. Chung, S.-L. Wu and Y.-K. Su, "Nonvolatile Bio-Memristor Fabricated with Egg Albumen Film," *Scientific Reports,* vol. 5, no. 1, 2015.

[12] Q. Sheng, Y. Xie, J. Li, X. Wang and J. Xue, "Transporting an ionic-liquid/water mixture in a conical nanochannel: a nanofluidic memristor," *Chem. Commun.,* vol. 53, no. 45, 2017.

[13] I. Köymen, P. A. Göktürk, C. Kocabaş and Ş. Süzer, "Chemically addressed switching measurements in graphene electrode memristive devices using in situ XPS," *Faraday Discuss.,* vol. 213, pp. 231-244, 2019.

[14] T. F. Schranghamer, A. Oberoi and S. Das, "Graphene memristive synapses for high precision neuromorphic computing," *Nature Communications,* vol. 11, no. 1, 2020.

[15] H. Abu Nahla, Y. Halawani, A. Alazzam and B. Mohammad, "NeuroMem: Analog Graphene-Based Resistive Memory for Artificial Neural Networks," *Scientific Reports,* vol. 10, no. 1, 2020.

[16] F. J. Romero, A. Toral, A. Medina-Rull, C. L. Moraila-Martinez, D. P. Morales, A. Ohata, A. Godoy, F. G. Ruiz and N. Rodriguez, "Resistive Switching in Graphene Oxide," *Front. Mater.,* vol. 7, 2020.

[17] J. Lee and W. D. Lu, ""On-Demand Reconfiguration of Nanomaterials: When Electronics Meets Ionics," *Advanced Materials,* vol. 30, no. 1, p. 1702770, 2018.

[18] Z. Wang, L. Wang, M. Nagai, L. Xie, M. Yi and W. Huang, "Nanoionics-Enabled Memristive Devices: Strategies and Materials for Neuromorphic Applications," *Advanced Electronic Materials,* vol. 3, no. 7, p. 1600510, 2017.

[19] M. V. Fedorov and R. M. Lynden-Bell, "Probing the neutral graphene–ionic liquid interface: insights from molecular dynamics simulations," *Phys. Chem. Chem. Phys.,* vol. 14, no. 8, p. 2552–2556, 2012.

[20] E. O. Polat and C. Kocabas, "Broadband Optical Modulators Based on Graphene Supercapacitors," *Nano Letters,* vol. 13, no. 12, pp. 5851-5857, 2013.

[21] N. Kakenov, O. Balci, T. Takan, O. V. A, H. Altan and C. Kocabas, "Observation of Gate-Tunable Coherent Perfect Absorption of Terahertz Radiation in Graphene," *ACS Photonics,* vol. 3, no. 9, p. 1531–153, 2016.

[22] S. Balci and C. Kocabas, "Graphene-Based Optical Modulators," in *Optical Properties of Graphene*, World Scientific, 2016, p. 435–456.

[23] S. H. Jo, T. Chang, I. Ebong, B. B. Bhadviya, P. Mazumder and W. Lu, "Nanoscale memristor device as synapse in neuromorphic systems," *Nano Letters,* vol. 10, no. 4, pp. 1297-1301, 2010.

[24] X. Zhu, D. Li, X. Liang and W. D. Lu, "Ionic modulation and ionic coupling effects in MoS2 devices for neuromorphic computing," *Nature Materials,* vol. 18, no. 2, pp. 141-148, 2019.

[25] M. Kettner, I. Vladimirov, A. J. Strudwick, M. G. Schwab and .. T. Weitz, "Ionic gel as gate dielectric for the easy characterization of graphene and polymer field-effect transistors and electrochemical resistance modification of graphene," *Journal of Applied Physics,* vol. 118, no. 2, p. 02550, 2015.

[26] F. Wang, M. E. Itkis, E. Bekyarova and R. C. Haddon, "Charge-compensated, semiconducting single-walled carbon nanotube thin film as an electrically configurable optical medium," *Nature Photonics,* vol. 7, pp. 459-465, 2013.

[27] V. K. Sangwan, H.-S. Lee, H. Bergeron, I. Balla, M. E. Beck, K.-S. Chen and M. C. Hersam, "Multi-terminal memtransistors from polycrystalline monolayer molybdenum disulfide," *Nature,* vol. 554, no. 7693, 2018.

[28] C. Wu, T. W. Kim, T. Guo, F. Li, D. U. Lee and J. J. Yang, "Mimicking Classical Conditioning Based on a Single Flexible Memristor," *Advanced Materials ,* vol. 29, no. 10, 2017.

[29] M. Copuroglu, P. Aydogan, E. Polat, C. Kocabas and S. Süzer, "Gate-Tunable Photoemission from Graphene Transistors," *Nano Letters,* vol. 14, no. 5, pp. 2837-2842, 2014.

[30] L. O. Chua, "Memristor- The missing circuit element," *IEEE Transactions on Circuit Theory,* vol. 18, no. 5, pp. 507-519, 1971.

[31] B. Linares-Barranco, T. Serrano-Gotarredona, L. Camuñas-Mesa, J. Perez-Carrasco, C. Zamarreño-Ramos and T. Masquelier, "On Spike-Timing-Dependent-Plasticity, Memristive Devices, and Building a Self-Learning Visual Cortex," *Frontiers in Neuroscience,* vol. 5, p. 26, 2011.